\documentclass[aps,prb,amsmath,amssymb,amsfont,showpacs,superscriptaddress,reprint]{revtex4-1}
\usepackage{amsmath}
\usepackage{xcolor}
\usepackage{graphicx}
\usepackage{times}
\usepackage{natbib}

\newcommand{\dd}{\mathrm{d}}
\DeclareMathAlphabet{\mathbfit}{OML}{cmm}{b}{it}

\DeclareMathOperator{\expe}{e}
\newcommand{\cplxi}{\mathrm{i}}

\begin{document}

\title{A time domain based method for the accurate measurement of Q-factor and resonance frequency of microwave resonators}


\author{B. Gy\"{u}re}
\affiliation{Budapest University of Technology and Economics, Department of Physics, Budafoki \'{u}t 8., Budapest, H-1111, Hungary}

\author{B. G. M\'{a}rkus}
\affiliation{Budapest University of Technology and Economics, Department of Physics, Budafoki \'{u}t 8., Budapest, H-1111, Hungary}

\author{B. Bern\'{a}th}
\affiliation{Budapest University of Technology and Economics, Department of Physics, Budafoki \'{u}t 8., Budapest, H-1111, Hungary}

\author{F. Mur\'{a}nyi}
\affiliation{Foundation for Research on Information Technologies in Society (IT'IS), Zeughausstrasse 43, 8004 Zurich, Switzerland}

\author{F. Simon}
\email[Corresponding author: ]{ferenc.simon@univie.ac.at}
\affiliation{Budapest University of Technology and Economics, Department of Physics, Budafoki \'{u}t 8., Budapest, H-1111, Hungary}

\date{\today}

\begin{abstract}
We present a novel method to determine the resonant frequency and quality factor of microwave resonators which is faster, more stable, and conceptually simpler than the yet existing techniques. The microwave resonator is irradiated at a frequency away from its resonance. It then emits an exponentially decaying radiation at its eigen-frequency when the excitation is rapidly switched off. The emission is down-converted with a microwave mixer, digitized and its Fourier transformation (FT) directly yields the resonance curve in a single shot. Being an FT based method, this technique possesses the Fellgett (multiplex) and Connes (accuracy) advantages and it conceptually mimics that of pulsed nuclear magnetic resonance. We also establish a novel benchmark to compare accuracy of the different approaches of microwave resonator measurements. This shows that the present method have similar accuracy to the existing ones.
\end{abstract}
\maketitle

Microwave resonators are employed in diverse branches of science and application. Examples for the earlier include microwave impedance measurements \cite{Gruner1,Gruner2,Gruner3}, particle accelerators \cite{mw_accelerator}, cosmic microwave studies, magnetic resonance spectroscopy and imaging \cite{PooleBook,EPRImaging}, and cavity quantum electrodynamics \cite{cavityQED}. Concerning applications, microwave resonators are used in e.g. communication as filters and source stabilizers \cite{LuitenFreqStab}, in microwave heating, and in radar sensing \cite{pozar2004microwave}. The microwave resonator is known to sustain microwave radiation in a form of a standing wave pattern at a resonance frequency, $f_0$ with a quality factor, $Q$ [Refs. \onlinecite{PooleBook,pozar2004microwave,ModeNote}]. The quality factor is related to the band-width (or FWHM) of the resonance curve, $\Delta f$, as $Q=f_0/\Delta f$. The resonance curve is a Lorentzian for single-mode resonators which reflects that the microwave field builds up and decays exponentially.

Characterization of $f_0$ and $Q$ is crucial for the microwave resonator applications. The existing methods are reviewed in Refs. \onlinecite{PetersanAnlage,LuitenReview}. The most common methods are to irradiate the resonator with a slowly varying frequency\cite{Gruner2} (analogue frequency sweep)  or at a few stabilized frequencies\cite{LuitenHiResQMeas} (stepped frequency sweep)  and to detect the reflection from or transmission through the resonator with a power detector. The $f_0$ and $Q$ are determined from the center of the Lorentzian and its width, respectively. The disadvantages of these methods are that i) they are prone to temporal instabilities as the curve is not measured at once, ii) analogue frequency sweep suffers from calibration and stability problems and stepped sweep is time consuming due to the finite frequency settling time of oscillators, iii) the reflection method suffers from the so-called standing wave problem \cite{PooleBook}, i.e. that imperfections of the microwave circuit distorts the Lorentzian curve.

Improved methods to measure $f_0$ and $Q$ include the use of a source whose frequency is stabilized to the resonator using the so-called automatic frequency control (AFC) methods (technically similar to the case of non-contact atomic force microscopy). Then, a varying $f_0$ is detected using a frequency counter \cite{Gruner2,Mehring}. However, absolute $Q$ values are available in these methods from an instrumental calibration only.

Time domain based methods were also developed \cite{LuitenReview} to study the microwave resonator parameters. The so-called \emph{decrement method} observes the transient resonator response when a microwave excitation, whose frequency is in resonance with the resonator, is switched on or off \cite{Gallagher,LuitenReview,EatonTransient}. Another variant is the so-called \emph{fast-sweep decrement method}, which monitors the resonator response to a rapidly swept microwave frequency \cite{SchmittZimmer,Kocherzhin}. The disadvantage of these methods is that i) \emph{a priori} knowledge of the resonance frequency is required and $f_0$ is not measured, ii) the $Q$ precision of these methods is known to be low \cite{LuitenReview}.

The present situation of microwave resonator measurement is similar to nuclear magnetic resonance (NMR) faced in its early years: a frequency or magnetic field swept NMR measurement suffered from poor accuracy of the resonance position and poor sensitivity due to the ineffective measurement technique. Pulsed NMR technique provides a simultaneous measurement at several frequencies, which is known as the multiplex or Fellgett advantage \cite{Fellgett} and a highly accurate value of the nuclear resonance (known as the Connes advantage \cite{Connes}), which led to an "NMR revolution" and the proliferation of high-resolution NMR\cite{Ernst}.

This compelling parallelism motivated us to adapt the pulsed NMR like methods to the measurement of $f_0$ and $Q$. We present a time-resolved technique, which allows their highly accurate and rapid measurement. The method is based on the observation of the transient microwave signal which occurs when the exciting power is rapidly switched off. The excitation can have an arbitrary frequency with some restrictions but not necessarily matching $f_0$. The transient signal is down-converted to the few kHz-MHz range with a microwave mixer where it is digitized and Fourier transformed, which yields the resonance curve. The $f_0$ and $Q$ are determined by fitting. The method exhibits multiplex advantage and the time domain data acquisition of a low frequency data allows the accurate calibration of the microwave frequencies and it is also free from the standing wave problem. It is well suited to study dynamic phenomena e.g. when studying laser induced photoconductivity in semiconductors with microwave resonators \cite{Subramanian}.



Principles of microwave resonators (mwc) are described in the literature \cite{PooleBook,pozar2004microwave} and we focus on the transient behavior, which occurs when the exciting microwave is switched on or off instantaneously. There is only one theoretical treatment of the microwave transient (Ref. \onlinecite{Gallagher}). Even that report is not considering the effect of irradiation frequency and it contains some errors. We give a rigorous treatment of resonator transients in Ref. \onlinecite{SupMat}.

For steady-state conditions, the frequency of the microwaves sustained inside the mwc matches that of the irradiation frequency, $f$, and the accumulated energy follows a Lorentzian profile as a function of $f-f_0$ with FWHM of $\Delta f$. The ratio of the reflected to transmitted power is described by the reflection coefficient, $\Gamma$,  which is 0 when the resonator is critically coupled and $f=f_0$ [Ref. \onlinecite{pozar2004microwave}]. Technically, the overall resonator $Q$ factor is measured, which reads: $Q^{-1}=Q_0^{-1}+Q^{-1}_{\text{c}}$, where $Q_0$ is the $Q$ factor of the unloaded (uncoupled) resonator and $Q_{\text{c}}$ is the $Q$ factor due to the resonator coupling. For critical coupling, $Q_{\text{c}}=Q_0$ and the lack of reflection for $f=f_0$ is an interference effect: the mwc continuously radiates through the coupling element, with a phase which is opposite to that reflected from the coupling element itself.

 The resonator transient behavior can be mathematically understood\cite{Gallagher} as a sum of the eigen- and driven oscillations. The earlier describes the transient and the latter the steady-state solution. When the exciting microwaves are switched off, the resonator starts to radiate \emph{instantaneously} with its eigenfrequency \cite{TransientNote}. It is somewhat more surprising but the resonator also emits radiation at its eigen-frequency for the switch on transient \cite{Gallagher,SchmittZimmer,Kocherzhin,EatonTransient}: the increasing microwave field inside the mwc during switch on is mathematically described by the sum of the decreasing negative amplitude transient (with frequency $f_0$) and the constant amplitude steady-state solution (with frequency $f$). Both components emanate from the resonator but the latter term is canceled due to the above described interference effect. The switch on and off transient signals have the same amplitude, decay constant, frequency of $f_0$, but opposite microwave phase, which can be readily detected with a phase sensitive microwave mixer whose local-oscillator (LO) port is driven with a microwave at $f$.

 The transient signal reads \cite{TransientNote2,SupMat} for the microwave power, $p(t)$, and  microwave voltage, $V(t)$:

 \begin{gather}
 p\left(t\right)=p_0\times \exp\left( -\frac{t\omega_0}{Q}\right)\\
 V(t) = \sqrt{p_0 Z_0}\times\exp\left( -\frac{t\omega_0}{2Q}\right)\times \exp\left(i\omega_0 t\right),
 \label{Eq:Transient}
 \end{gather}

\noindent where $p_0$ is the power of the source, $\omega_0=2\pi f_0$, $Z_0$ is the wave impedance of the microwave waveguide and we omitted the phase of the reflected microwaves in $V(t)$. The Fourier transformation (FT) of $V(t)$ yields a Lorentzian peaked at $f_0$ with FWHM of $\Delta f=f_0/Q$.

\begin{figure}[htp]
\begin{center}
\includegraphics[width=0.45\textwidth]{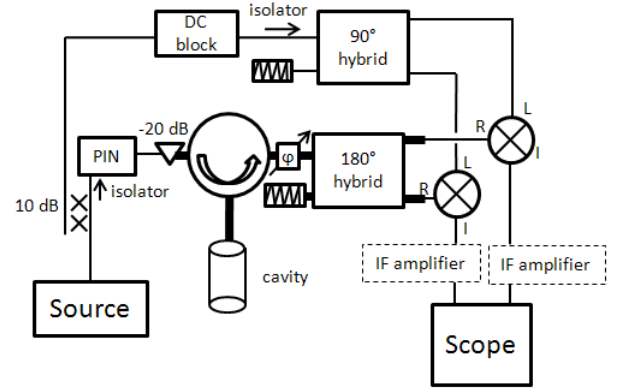}
\caption{Schematics of the instrument used to detect the microwave resonator transients. Thin and thick lines represent coaxial and microwave waveguides, respectively. The circulator, $180\,\deg$ hybrid (or Magic Tee) are also waveguide elements.}
\label{Fig:TimeDom_schem}
\end{center}
\end{figure}

Fig. \ref{Fig:TimeDom_schem}. shows the setup to detect the microwave resonator transients. The microwave source is a Gunn diode stabilized to a quartz oscillator with a PLL system and has 20 dBm leveled output power (Agilent 83751B, 2-20 GHz). The coupled port of a 10 dB directional coupler serves as LO mixer input. Microwaves are switched with a fast PIN diode with on-off transition time less than 5 ns (Advanced Technical Materials, S1517D). The PIN diode is driven by an arbitrary waveform generator (HP33120A) with varying pulse length and frequency. The microwaves are attenuated to a level of 0 dBm (1 mW) and a coaxial to microwave waveguide transition serves as band filter since the PIN diode output contains crosstalk from the driving signal. The microwaves are directed towards and from the resonator using a waveguide circulator. We studied two TE011 cylindrical cavities with a variable lateral iris coupling \cite{PooleBook} of copper ($Q_0\approx 10^{4}$) at room temperature and niobium kept at 4.2 K ($Q_0\approx 10^{5}$) with equal length and diameter of about 3 cm.

The reflected microwave is phase shifted, its power is split with a waveguide magic Tee and it serves as the RF mixer input with a maximum input power of 0 dBm to avoid mixer saturation. The LO port of the mixers is protected by an inside/outside DC block against crosstalk from the PIN diode driving signal. The 10 dBm LO power is split on a $90 ^\circ$ hybrid to provide the LO inputs of 7 dBm for the mixers (Marki Microwave M10418LC, NF=6 dB, IF=DC-4 GHz, RF/LO=4-18 GHz). The two mixer configuration functions as a quadrature or I/Q mixer. The intermediate frequency (IF) signal of the mixers is optionally amplified (Analogue Modules 322-6-50, low noise voltage amplifier). The IF signal is digitized by a 500 MHz oscilloscope (LeCroy LT342) which also performs signal processing (averaging and FT). Automation is achieved by computer control of the source, PIN diode driver, the oscilloscope read-out, and data analysis.

\begin{figure}[htp]
\begin{center}
\includegraphics[width=0.45\textwidth]{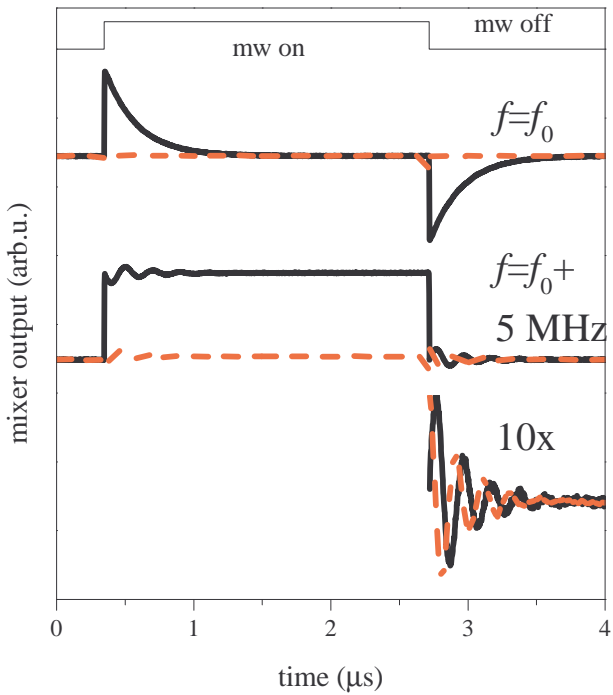}
\caption{Microwave transient for a critically coupled resonator when $f=f_0$ and $f=f_0+5\,\text{MHz}$. Solid and dashed curves are the in- and out-of-phase mixer outputs; the latter is nulled for $f=f_0$. Incident power is 0 dBm, noise is not visible. Note the different sign of the transient for switch on and off and the damped oscillations when $f \neq f_0$. The latter is also shown on a 10x scale.}
\label{Fig:Transient_measurements}
\end{center}
\end{figure}

The measured transient signals are shown in Fig. \ref{Fig:Transient_measurements} for on-resonance ($f=f_0$) and off-resonance ($f \neq f_0$) excitations and critical coupling. The microwave phase was adjusted to null the signal on one of the mixers. We obtained qualitatively similar data for under- and over-coupling. Exponential decays are observed for both switch on and off with opposite signs when $f=f_0$, which is in agreement with the expectation. After switch on, the reflected microwave field decays to zero due to critical coupling. An exponentially decaying signal, oscillating with $f-f_0$ frequency, is observed for both mixers when $f\neq f_0$. This fully supports the above qualitative arguments that the resonator radiates microwaves at its eigenfrequency for both switch on and off.

The switch off transient signal is analogous to the \textit{free induction decay} of NMR spectroscopy. We note that the exponentially decaying transient was observed for $f=f_0$ using a power detector in Refs. \onlinecite{Gallagher,EatonTransient} but no systematic study of this phenomenon neither its applications were pursued. Our instrument represents advances in the following respects: i) we use phase sensitive mixer detection rather than the previous phase independent power diode measurements, ii) and we are not restricted to the $f=f_0$ case, and iii) we analyze the data with FT to yield the resonator parameters quantitatively.

The quadrature detection allows to perform a complex FT of the transient signal it thus provides the sign of $f-f_0$. The FT power spectrum is a Lorentzian with FWHM of $\Delta f$ according to Eq. \ref{Eq:Transient}. Although information content is identical in the two transients (switch on or off), in practice we use the switched off one as i) it allows the use of larger microwave powers without mixer saturation, ii) the FT is less affected by the DC background or the well known zero-frequency anomaly, and iii) microwave standing waves are absent. We also considered an alternative: the switch off transient after the application of a short pulse in an even closer analogy to NMR but it results is smaller signals. In practice, it is best to irradiate the resonator until the switch on transient decays and to observe the subsequent switch off transient.

\begin{figure}[htp]
\begin{center}
\includegraphics[width=0.45\textwidth]{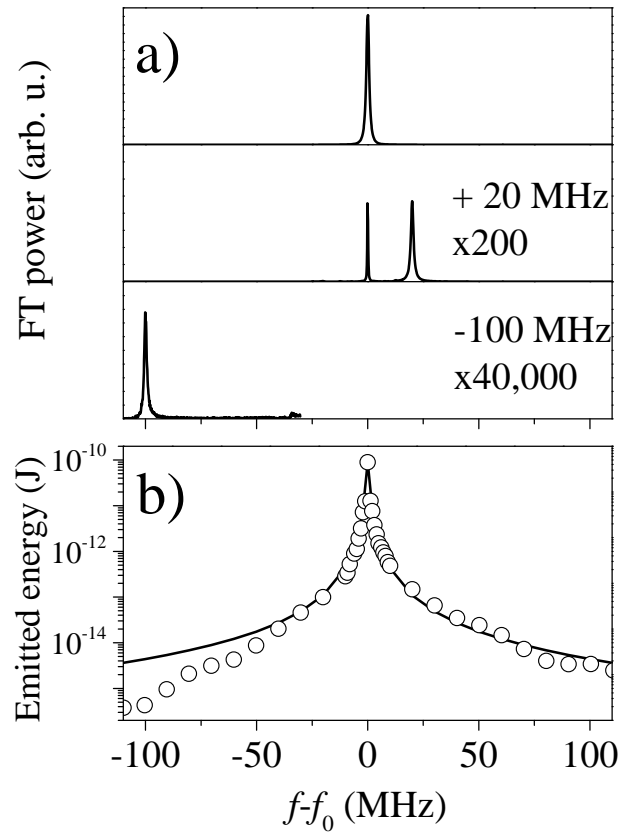}
\caption{(a) FT data for different values of $f-f_0$ (from top to bottom): 0, +20 MHz, and -100 MHz. Note the different vertical scales, the DC peak for the +20 MHz data, and that this range is not shown for the -100 MHz data. (b) The energy emitted from the resonator, obtained from the integrated intensity of the FT signal, as a function of $f-f_0$. The calculated solid curve is explained in the text.}
\label{Fig:FT_detuning}
\end{center}
\end{figure}

Fig. \ref{Fig:FT_detuning}a. shows the FT of the switch off transient data as a function of the resonator detuning, $f-f_0$. A fit yields both $f_0$ and the width of the Lorentzian peak, which provides $Q$. The accuracy of these parameters is discussed below. The FT data shows a DC peak due to some offset in the mixer output. There is a small quadrature ghost at the image frequency of the FT due to a small, less than $5\,^\circ$ error of the $90\,^\circ$ hybrid.

It may appear surprising that microwave emission is observed from a resonator with $\text{FWHM}\approx 1\,\text{MHz}$ when the exciting source is detuned up to 100 MHz from $f_0$. Fig. \ref{Fig:FT_detuning}b. shows the measured energy which is emitted from the resonator and is obtained from the integral of the FT power spectrum. We considered the losses in the microwave circuit, the mixer conversion loss, and the mixer wave impedance ($50\,\Omega$) to obtain energy from the voltage output of the mixer. A vertically scaled calculated curve for the emitted energy, $U_{\text{em}}$, is also shown in Fig. \ref{Fig:FT_detuning}b. according to Ref. \onlinecite{SupMat}:

\begin{gather}
U_{\text{em}}=\frac{p_0 Q_0}{2 \omega_0}\times\frac{\left(\frac{\Delta \omega}{2}\right)^2}{\left(\frac{\Delta \omega}{2}\right)^2+\left(\omega-\omega_0\right)^2}.
\label{Eq:EmittedEnergy}
\end{gather}

\noindent The measured and calculated emitted energy data are in good agreement except for $f-f_0<-50\,\text{MHz}$; the latter is probably due to a reduced performance of the microwave circuit over this relatively large band-width. Nevertheless, this agreement attests that our theoretical description of the resonator energy is appropriate.

 Given the equivalent noise band-width of the measurement (ENBW) in Hz units, the limit of detection is typically $10^{-9}\,\text{J}\times \text{ENBW}$ for a power detector and $10^{-20}\,\text{J}\times \text{ENBW}$ for a mixer \cite{pozar2004microwave}. Clearly, a power detector is not capable of studying the case of significant detuning, however for our resonator of $\text{FWHM}\approx 1\,\text{MHz}$ and ENBW=100 kHz, one can study it up to 100 MHz detuning and even larger values up to 1 GHz could be attained with increased exciting power\cite{SupMat}.


We compare the accuracy of our method with those available in the literature. There exists no common standards as to how the different resonator characterization techniques could be compared. While all methods have a preferred range of $Q$, we expect that a proper 'figure of merit of the measurement technique' is $Q$ independent.

We denote the standard deviation and mean of the corresponding quantity with $\sigma(\cdot)$ and $\overline{\cdot}$, respectively. Alternatives in the literature to express the accuracy of the measurement are $\sigma\left(1/2Q\right)$ [Ref. \onlinecite{Mehring}] and $\sigma \left(Q\right)/\overline{Q}$ [Ref. \onlinecite{LuitenReview}] for $Q$ and
$\sigma \left(f_0\right)/\overline{f_0}$ [Ref. \onlinecite{LuitenHiResQMeas}] for $f_0$. However, of these $\sigma \left(1/2Q\right)$ and $\sigma\left(f_0\right)/\overline{f_0}$ are not appropriate as these change if $Q$ changes.

We define the error of the $Q$ and $f_0$ measurement as:

\begin{align}
\delta\left(Q\right) := \frac{\sigma \left(Q\right)}{\overline{Q}}\,, \delta\left(f_0\right):=\frac{\sigma \left(f_0\right)}{\overline{\Delta f}},
\label{Eq:ErrorDefinition}
\end{align}

\noindent which in turn do not change if the $Q$ factor changes and the accuracy of the method remains the same.
To highlight the merit of these error definitions, we give the corresponding values for two literature methods, which has proven to be the most accurate. The AFC based method \cite{Mehring} gives $\delta\left(Q\right)=10^{-3}$  for a 3 sec measurement (from the quoted values of $Q=25.000$ and $\sigma\left(1/2Q\right)=10^{-8}$). The stabilized stepped frequency method \cite{LuitenHiResQMeas} gives $\delta\left(Q\right)=6\times 10^{-4}$ for a $Q \approx 10^9$ resonator (10 sec measurement). We find it reassuring that two different techniques for two very different $Q$ values provide very similar $\delta\left(Q\right)$ values. We do not have a consistent explanation, why every technique \cite{PetersanAnlage,LuitenReview} (including ours) converge to this limit of $\delta\left(Q\right)\approx 10^{-3}$ but it hints at a common technical limit. We recommend to use Eq. \eqref{Eq:ErrorDefinition}. as a standard benchmark to characterize the resonator parameter measurement accuracy with the error normalized to 1 second measurement time.

Another observation is that $\delta\left(Q\right)\approx\delta\left(f_0\right)$ holds for the values in Refs. \onlinecite{LuitenHiResQMeas,Mehring} and for all kinds of resonator measurements which we tested (frequency sweep method, AFC method, and the present method). A calculation of the error propagation yields:

\begin{gather}
\delta\left(Q\right)\approx\frac{\sigma\left(\Delta f\right)}{\overline{\Delta f}}\approx \frac{\sigma \left(f_0\right)}{\overline{\Delta f}}.
\label{Eq:SimplifiedError}
\end{gather}

\noindent Given that $f_0 \gg {\overline{\Delta f}}$, Eq. \eqref{Eq:SimplifiedError} is equivalent to $\sigma\left(\Delta f\right)\approx \sigma\left(f_0\right)$, which is reasonable given that both parameters are determined from the same data. This observation underlines the value of this definition, Eq. \eqref{Eq:ErrorDefinition}, as it provides the \emph{same} error value for both resonator parameters.

\begin{table}[htp]
\begin{center}
    \begin{tabular*}{0.45\textwidth}{@{\extracolsep{\fill}}lllll}
    \hline \hline
    Method & $Q$ & $t\,\text{[sec]}$ & $\delta\left(Q\right)$ & $\delta\left(f_0\right)$\\ \hline
    Ref. \onlinecite{LuitenHiResQMeas} & $10^{8}-10^{9}$ & 10 & $6\times 10^{-4}$ & $6\times 10^{-4}$ \\
    Ref. \onlinecite{Mehring} & $2.5\times 10^{4}$ & 3 & $10^{-3}$ & $10^{-3}$\\
    Present method & $10^4-10^5$ & 1 & $10^{-3}$ & $10^{-3}$\\
    \hline \hline
    \end{tabular*}
    \caption{Comparison of the different $Q$ and $f_0$ measurement methods ($t$ denotes the measurement time).}
    \label{tab:Comparison}
\end{center}
\end{table}

In Table. \ref{tab:Comparison}., we compare the errors of the present and the two best performing literature methods (Refs. \onlinecite{LuitenHiResQMeas,Mehring}). We find that the present method has similar error for a similar measurement time. The present method is limited for $Q>1000$ values due to the available time resolution and PIN diode switching speed. However, we expect it to perform better than the conventional methods for higher $Q$ values (even up to the $Q=10^9$ range) as therein the limiting factors are the measurement speed where the present method with its multiplex advantage is a true asset. We remind that this is very similar to the multiplex advantage which motivated the development of the pulsed NMR technique. When this parallel is followed, we expect that phase-cycling like methods (which are standard in NMR) could further improve the accuracy of the present technique and to remove some of the spurious electronic response (e.g. mixer DC offset). The absolute accuracy of both and $f_0$ and $Q$ is traced back to the accuracy of the local oscillator, which can be very high with the use of atomic clocks referencing. This is essentially the so-called Connes (accuracy) advantage \cite{Connes} of the FT based techniques.

Another advantage of the present method is the availability to measure dynamics of microwave absorption inside microwave resonators. Effects like sample heating \cite{KarsaPSSB} are known to affect the microwave resonator parameters during a measurement. The conventional methods are limited to few ms-s measurement time, whereas for the present one, the only time limit is the resonator transient time itself.

In conclusion, we presented a novel method to measure the parameters of microwave resonators. It is based on the phase sensitive observation of transient signals which arise when the resonator microwave excitation is switched on or off. No prior knowledge of the resonant frequency is required and the method is superior to existing alternatives in terms of stability, measurement speed, and conceptual simplicity. We proposed a novel benchmark to evaluate the figure of merit of different resonator measurement methods and we find that accuracy of the present method is comparable to the known alternatives.

Work supported by the European Research Council ERC-259374-Sylo Grant. The authors are indebted to F. I. B. (Tito) Williams, Andr\'{a}s J\'{a}nossy, and K\'{a}roly Holczer for stimulating discussions.



\bibliographystyle{apsrev}
\bibliography{Gyure_etal_condmat_submit}

\begin{thebibliography}{27}
\bibitem[{A single- or mono-mode cavity sustains radiation for a given
  frequency and its harmonics, whereas a multi-mode cavity can sustain
  radiation for a number of different frequencies with dissimilar standing wave
  patterns()}]{ModeNote}
A single- or mono-mode cavity sustains radiation for a given frequency and its
  harmonics, whereas a multi-mode cavity can sustain radiation for a number of
  different frequencies with dissimilar standing wave patterns.
\bibitem[{This is a shere mathematical consequence of the driven oscillator
  equation of the cavity: eigen-oscillation starts without any memory of the
  previous frequency. The initial conditions only determine the oscillation
  amplitude and phase.()}]{TransientNote}
This is a shere mathematical consequence of the driven oscillator equation of
  the cavity: eigen-oscillation starts without any memory of the previous
  frequency. The initial conditions only determine the oscillation amplitude
  and phase.
\bibitem[{Note that our result differs from Eq. (36) in Ref.
  \onlinecite{Gallagher} with a factor 2.()}]{TransientNote2}
Note that our result differs from Eq. (36) in Ref. \onlinecite{Gallagher} with
  a factor 2.
\end{thebibliography}

\pagebreak
\begin{widetext}

\section*{Supplementary information}
\appendix

This supplementary material is organized as follows. We first discuss microwave cavity parameters in general terms including the physics of coupling and reflected microwaves. We then discuss the switch on and off transients for on-resonance ($f=f_0$) conditions and the case of arbitrary coupling using a conservation of energy argument. This is used to deduce Eqs. (1) and (2) of the main manuscript. An empirical verification of the result is also given.

The results on the cavity transients is generalized for $f \neq f_0$ using a circuit model. We note that to our knowledge this derivation is not available elsewhere. We derive Eq. (3) in the manuscript and provide the limit of transient detection for the case of significant detuning ($\left|f-f_0\right|\gg \Delta \omega$). It provides an estimate of the maximum available detuning while the cavity transient is observable for a given cavity excitation power. We finally provide additional cavity transient data for an undercoupled cavity and a superconducting cavity with higher quality factor.

\section{The cavity parameters}

There is some disagreement even amongst seminal contributions about what is meant by the cavity $Q$ or quality factor. The cavity resonance frequency, $f_0$, is agreed to be the frequency where the largest energy is stored in the cavity.

To clear the issue of the quality factor, we first consider a microwave cavity which is weakly coupled to its environment. It means that much of the incoming power, $p_0$, is reflected from it. We define the exciting power, $p_{\text{exc}}$, which is the power which enters the cavity and excites its microwave field and it holds: $p_{\text{exc}}\ll p_0$. Then, the reflected microwave power follows a downward pointing Lorentzian curve on top of a constant background as a function of the exciting frequency, $f$, which is centered at $f_0$ and has a FWHM of $\Delta f$. From the measurable quantities we can define the \textit{quality factor of the undercoupled cavity}, $Q_0$ as
\begin{equation}
Q_0 = \frac{f_0}{\Delta f} = \frac{\omega_0}{\Delta \omega_0},
\label{eq:Q0_resonancedef}
\end{equation}
where $\omega_0=2\pi f_0$. The other common definition is coming from the stored energy
\begin{equation}
Q_0 = \omega_0 \frac{U}{p_{\text{exc}}},
\label{eq:Q0_energydef}
\end{equation}
where $U$ is the total stored energy in the cavity. We shall show later that the above two definitions in Eq. \eqref{eq:Q0_resonancedef} and Eq. \eqref{eq:Q0_energydef} are equivalent. When we consider the conservation of energy inside the cavity, the dissipated power, $p_{\text{diss}}$ equals $p_{\text{exc}}$. If we rearrange Eq. \eqref{eq:Q0_energydef}, we obtain that
\begin{equation}
p_{\text{diss}}=\frac{U\omega_0}{Q_0}.
\label{eq:Q0_dissipationdef}
\end{equation}
The role of $Q_0$ in Eq. \eqref{eq:Q0_dissipationdef} can be considered as a proportionality constant between the dissipated power and the stored energy. Physically, the dissipation occurs due to eddy currents in the cavity and the loss only depends on the cavity material. It means that the proportionality between the loss and the stored energy remains the same irrespective of the level of coupling of the cavity to the waveguide.

When the coupling of the cavity to the waveguide is increased, the amount of exciting power increases until it reaches $p_0$ for critical coupling, i.e. when there is no reflected power from the cavity. At the same time, the reflected power profile as a function of $f$ broadens and it is twice as broad for critical coupling than for the undercoupled case. In the following, we use $Q$ for the \textit{quality factor of the coupled cavity}. Technically, it is the coupled $Q$ factor which is observable and $Q_0$ can only be approximately measured when the undercoupled cavity is studied or it can be deduced by indirect means from $Q$.

The coupling element is most commonly an iris \cite{PooleBook} which reflects most of the incoming power and transmits only a fraction of it. It means that the iris itself does not contribute to any dissipation or loss, still the nominal $Q$ factor of the coupled cavity is expressed using the \textit{quality factor of the coupling}, $Q_{\text{c}}$, as:
\begin{equation}
Q^{-1}=Q_0^{-1}+Q_{\text{c}}^{-1}.
\label{eq:Q_vs_Qc}
\end{equation}

This also implies that Eq. \eqref{eq:Q0_energydef} does not hold for the coupled $Q$, only for the undercoupled value. Eq. \eqref{eq:Q_vs_Qc} explains that for critical coupling $Q=Q_0/2$ as therein $Q_{\text{c}}=Q_0$. Clearly, the critical coupling is a distinguished case. We show below that this is not only due to the lack of power reflection but it also means that power dissipated inside the cavity equals the power transmitted through the coupling element.

\begin{figure}[h!]
\begin{center}
\includegraphics[width=0.500\textwidth]{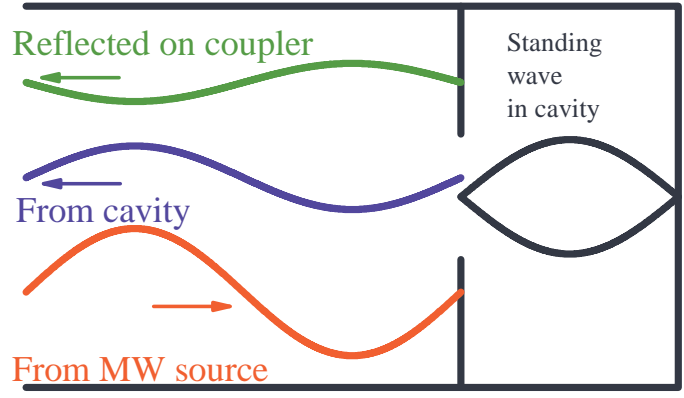}
\caption{Schematics of the iris of a cavity and the reflected, transmitted, and radiation emanating from the cavity.}
\label{fig:iris}
\end{center}
\end{figure}

Fig. \ref{fig:iris}. shows the schematics of the microwave cavity and the coupling element. The coupling element is an iris in the figure which reflects $R$ amount of the incoming power and transmits $T$ portion of it and $R+T=1$. There are three microwave components which are to be considered: the microwave field that is reflected from the iris, $E_{\text{refl}}$, the microwave field which enters the cavity and excites it, $E_{\text{exc}}$, and the microwave field that emanates from the cavity through the iris, $E_{\text{em}}$. It is the interference between the first and third terms which results in zero reflected power from the cavity for critical coupling and irradiation for $f=f_0$ as these have opposite microwave phase \cite{PooleBook}. Often, this interference effect is referred to as \emph{microwave field reflected from the cavity}. The thorough description of this interference effect is important for the cavity transient signals as the amount of reflected power depends on the state of the cavity \cite{Gallagher}.

Conservation of energy dictates that then $p_{\text{exc}}=p_0=p_{\text{diss}}$. The energy accumulated inside the cavity is then $U=p_0 Q_0/\omega_0$, thus the emanated electric field is:
\begin{equation}
E_{\text{em}}=\sqrt{U \omega_0 T},
\label{eq:emanating_field}
\end{equation}
whose magnitude equals that of the reflected electric field:
\begin{equation}
E_{\text{refl}}=\sqrt{p_0 R}.
\label{eq:reflected_field}
\end{equation}

Eqs. \eqref{eq:emanating_field} and \eqref{eq:reflected_field} yield that
\begin{gather}
R = \frac{Q_0}{1+Q_0}, \\
T = \frac{1}{1+Q_0}.
\label{eq:t}
\end{gather}

\section{The cavity transients}

A preliminary note is required concerning the nature of transients. In the main paper, we refer to the signal which is measured by the microwave mixers as \emph{cavity transient signals}. For the switch off transient, the signal comes from the discharging cavity, whereas for the switch on transient, the signal is a reflection from the charging cavity.

\subsection{The switch off transient}

We first consider a critically coupled cavity which is irradiated at its resonance frequency, $f=f_0$, in its stationary state with energy $U(t=0)=U_0=p_0 Q_0/\omega_0$. The cavity loses energy through two paths after the excitation is switched off: by radiation through the coupling element and by dissipation and the corresponding differential equation for the stored energy $U(t)$ reads:
\begin{equation}
\frac{\dd U}{\dd t} = -p_{\text{em}} - p_{\text{diss}} = - U \omega_0 T - \frac{U \omega_0}{Q_0}.
\end{equation}

Which yields after rearranging:
\begin{equation}\label{eq:losses}
\frac{\dd U}{\dd t} = - U \omega_0\left( \frac{1}{1+Q_0} + \frac{1}{Q_0} \right)=-\frac{U \omega_0}{Q}.
\end{equation}
Where the quality factor of the coupled cavity is:
\begin{equation}\label{eq:qapprox}
\frac{1}{Q} = \frac{2Q_0 + 1}{(1 + Q_0)Q_0} \approx \frac{2}{Q_0}.
\end{equation}
The approximation is valid when $Q_0 \gg 1$, which is often satisfied as $Q$ values in excess of $1.000$ are customary. The solution for the critically coupled case and $f=f_0$ thus reads:
\begin{equation}
U(t) = U_0 \expe^{-\frac{\omega_0 t}{Q}} \approx U_0 \expe^{-\frac{2\omega_0 t}{Q_0}}.
\end{equation}

Using that $p_{\text{em}} = U \omega_0 T$ we obtain Eqs. (1) and (2) of the main paper for the power and the amplitude of the microwave voltage which is emitted from the cavity and read:
\begin{gather}
p_{\text{em}}(t)=p_0 \expe^{-\frac{\omega_0 t}{Q}} \approx p_0 \expe^{-\frac{2\omega_0 t}{Q_0}}, \\
V_{\text{em}}(t)=\sqrt{p_0 Z_0} \expe^{-\frac{\omega_0 t}{2Q}} \approx \sqrt{p_0 Z_0} \expe^{-\frac{\omega_0 t}{Q_0}},
\label{eq:transient_equations}
\end{gather}
where $Z_0$ is the wave impedance of the of the microwave waveguide.

\subsection{The switch on transient}

After switch on the energy balance of the cavity reads (for critical coupling and $f=f_0$):
\begin{equation}
\frac{\dd U}{\dd t}=p_{\text{exc}}-p_{\text{diss}},
\label{eq:switch_on_trans_simple}
\end{equation}
where $p_{\text{diss}}=U\frac{\omega_0}{Q_0}$ due to the arguments above. During the transient, the power reflected from the cavity is not zero and thus the exciting power reads:
\begin{equation}
p_{\text{exc}}=p_0-\left|E_{\text{refl}}-E_{\text{em}} \right|^2,
\end{equation}
which equals according to Eqs. \eqref{eq:emanating_field} and \eqref{eq:reflected_field}:
\begin{equation}
p_{\text{exc}}=p_0-\left|\sqrt{p_0 R}-\sqrt{U\omega_0 T} \right|^2.
\end{equation}

Eq. \eqref{eq:switch_on_trans_simple} reads when $Q_0\gg 1$:
\begin{equation}
\frac{\dd U}{\dd t} = 2\sqrt{\frac{p_0 U \omega_0}{Q_0}} - 2\frac{U\omega_0}{Q_0}.
\end{equation}

It can be readily verified that the solution
\begin{equation}
U(t)=\frac{p_0 Q_0}{\omega_0}\left(1-\expe^{-\frac{\omega_0 t}{Q_0}}\right)^2,
\end{equation}
satisfies the starting conditions and the differential equation. The power which is reflected from the cavity during the transient is obtained as:
\begin{equation}
\left|\sqrt{p_0}-\sqrt{\frac{U \omega_0}{Q_0}} \right|^2=p_0 \expe^{-\frac{2\omega_0 t}{Q_0}},
\end{equation}
and the corresponding amplitude of the microwave voltage reads: $\sqrt{p_0 Z_0}\expe^{-\frac{\omega_0 t}{Q_0}}$. These results are identical to that in Eq. \eqref{eq:transient_equations}. It also confirms that Eqs. (1) and (2) of the main paper are valid for both the switch on and off transients.

Finally, we note that often a relaxation time is defined to describe the cavity transient \cite{PooleBook}. It is however misleading as one has to specify whether the relaxation time is referred for the microwave voltage or microwave power.

\subsection{Empirical verification of Eqs. (1) and (2) of the main paper}

The cavity transient signals are measured with a microwave mixer, i.e. the amplitude of the microwave voltage is measured. A unitary Fourier transformation for $t=[0,\infty)$ of $V(t)$ in Eq. \eqref{eq:transient_equations} gives $V(\omega)$:
\begin{equation}
V(\omega)=\frac{\sqrt{p_0 Z_0}}{\sqrt{2\pi}}\times \frac{\frac{\Delta \omega}{2} +\cplxi \omega}{\left(\frac{\Delta \omega}{2}\right)^2+\omega^2},
\end{equation}
where we introduced $\Delta \omega=\omega_0/Q$. The power of the Fourier transformed $V(t)$ is $\left| V(\omega)\right|^2$ and reads:
\begin{equation}
\left|V(\omega)\right|^2=\frac{p_0 Z_0}{2\pi} \times \frac{1}{\left(\frac{\Delta\omega}{2}\right)^2+\omega^2}.
\label{eq:power_spectrum}
\end{equation}

Eq. \eqref{eq:power_spectrum} describes a Lorentzian curve with FWHM of $\Delta \omega$. This is in fact the generally accepted definition of the quality factor for a critically coupled cavity, i.e. that $Q$ is the ratio between $\omega_0$ and the FWHM of the resonance profile. In the following, we present our measurements of the Fourier transformed cavity transient power spectrum and compare it with the more conventional frequency swept data. The latter is recorded with a microwave power detector.

\begin{figure}[h!]
\begin{center}
\includegraphics[width=0.500\textwidth]{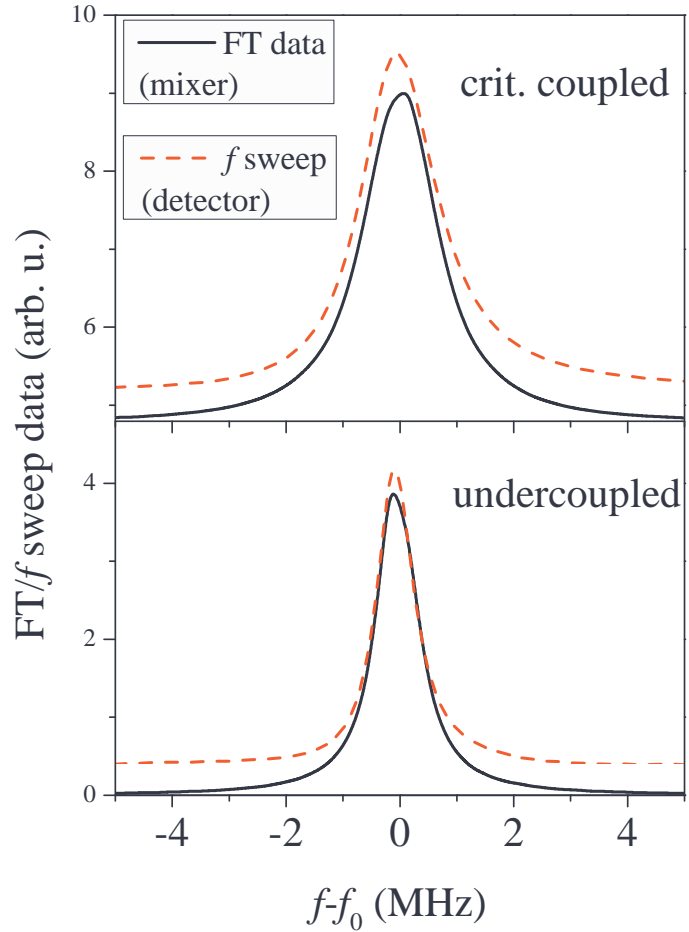}
\caption{Comparison of the FT power spectrum of the cavity transient measured with a mixer with that obtained with a power detector while the irradiation frequency was swept. Note that there is no scaling for the frequency axis, except that the 0 was offset to the cavity resonance frequency. The vertical axis is scaled however.}
\label{fig:FT_transient_Freq_swept}
\end{center}
\end{figure}

In Fig. \ref{fig:FT_transient_Freq_swept}. we compare the Fourier transform power spectrum of the cavity transient measured with microwave mixers with the data obtained with power detectors while the irradiation frequency was swept. The data is shown for two different setups, critically coupled and undercoupled case, with $Q=6400$ and $Q=12.800$, respectively. The data clearly shows that the two kinds of measurement produce equivalent results for both critical and undercoupled cases. This shows that the measurement using the cavity transient with mixers is indeed an appropriate way of measuring the cavity resonance profile.

We note an interesting consequence of Eq. \eqref{eq:power_spectrum}. When $\left|V(\omega)\right|^2/Z_0$ is integrated according to $\omega$, its dimension is Joule and its value is:
\begin{equation}
\int\limits_{-\infty}^{\infty} \frac{\left|V(\omega)\right|^2}{Z_0} \, \dd \omega=\frac{p_0}{\Delta \omega}=\frac{p_0 Q_0}{2 \omega_0},
\end{equation}
where we used that the FWHM of the coupled cavity is $\Delta \omega=2\Delta \omega_0=2\omega_0/Q_0$. This is \emph{half} of the energy stored inside the cavity in the stationary case. It means that the total microwave energy reaching the mixer is half of the stored energy, which is in agreement with our above description that half of the stored energy is dissipated and the other half is radiated through the coupling element. This underlines that our description is self consistent.

\subsection{Transients for arbitrary coupling}

Herein, the case of non-critical coupling is considered for irradiation at resonance, i.e. $f=f_0$. The exciting power inside the cavity, $p_{\text{exc}}$ is not necessarily $p_0$, and there is a finite reflected power even in the steady state. To describe this situation, the $\beta=\frac{Q_0}{Q_{\text{c}}}$ factor is introduced. As a result, for arbitrary coupling the cavity $Q$ factor reads $Q=Q_0/(1+\beta)$. The power exciting the cavity is:
\begin{equation}
p_{\text{exc}}=p_0\frac{4 \beta}{\left(1+\beta \right)^2}
\end{equation}
and the power reflected from the cavity is $p_0-p_{\text{exc}}$. It can be readily shown using the conservation of energy that the microwave power leaving the cavity is $\beta p_{\text{exc}}$ as it satisfies the requirement:
$\left(\sqrt{p_0}-\sqrt{\beta p_{\text{exc}}}\right)^2=p_0-p_{\text{exc}}$.

The differential equation for the energy stored inside the cavity reads for the switch off transient:
\begin{equation}
\frac{\dd U}{\dd t} = - (1+\beta)p_{\text{exc}} = -(1+\beta) U \frac{\omega_0}{Q_0}.
\end{equation}
This is solved with the $U(t=0)=U_0=p_{\text{exc}}Q_0/\omega_0$ initial condition
\begin{equation}
U(t) = U_0 \expe^{-(1+\beta)\frac{\omega_0 t}{Q_0}} = U_0 \expe^{-\frac{\omega_0 t}{Q}}.
\end{equation}

The amplitude of the microwave voltage which is detected during the transient reads:
\begin{equation}\label{eq:discvolt}
V(t)=\sqrt{p_{\text{exc}} Z_0 \beta} \expe^{-\frac{\omega_0 t}{2Q}}=\frac{2\beta}{1+\beta}\sqrt{p_0 Z_0}\expe^{-\frac{\omega_0 t}{2Q}}.
\end{equation}

Considering the switch on transient yields the same result for the power and microwave voltage amplitudes which are reflected from the cavity. Clearly, the Fourier transform analysis of Eq. \eqref{eq:discvolt} yields a Lorentzian in frequency space which corresponds to the quality factor of $Q$ and it is therefore the generalization of Eqs. (1) and (2) in the main paper.

\subsection{The transients in an equivalent circuit model}

The previous considerations were valid for on-resonance, i.e. $f=f_0$, and the calculations were based on the conservation of energy and wave interference effects. For the general, $f \neq f_0$ case a lumped circuit equivalent of the iris coupled microwave cavity has to be considered.

The lumped circuit equivalent of an iris-coupled cavity according to Ref. \onlinecite{princmic} is shown in Fig. \ref{fig:rlc-iris_paral}. for arbitrary $Q$ and Fig. \ref{fig:rlc-iris}. The resonant part (i.e. the cavity) is described and RLC circuit and an inductive coupling element with inductance $\mathcal{L}$ is considered. For high-$Q$, $L \gg \mathcal{L}$ is satisfied.

\begin{figure}[h!]
\begin{minipage}[b]{0.47\textwidth}
\centering
\includegraphics[width=0.9\textwidth]{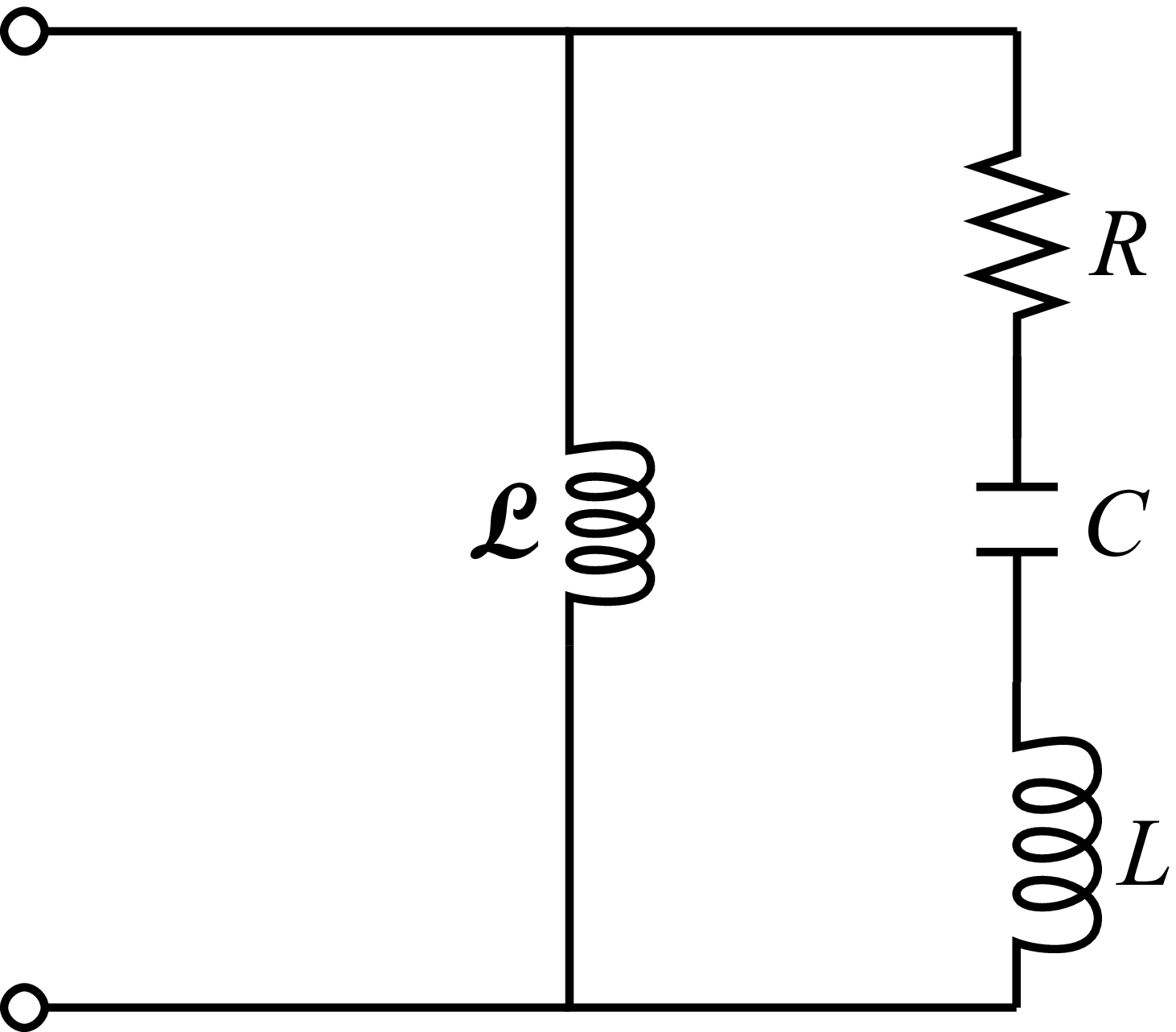}
\caption{Equivalent RLC-circuit of iris coupled cavity near resonance\cite{princmic}.}
\label{fig:rlc-iris_paral}
\end{minipage}
\hspace{0.75cm}
\begin{minipage}[b]{0.47\textwidth}
\centering
\includegraphics[width=0.9\textwidth]{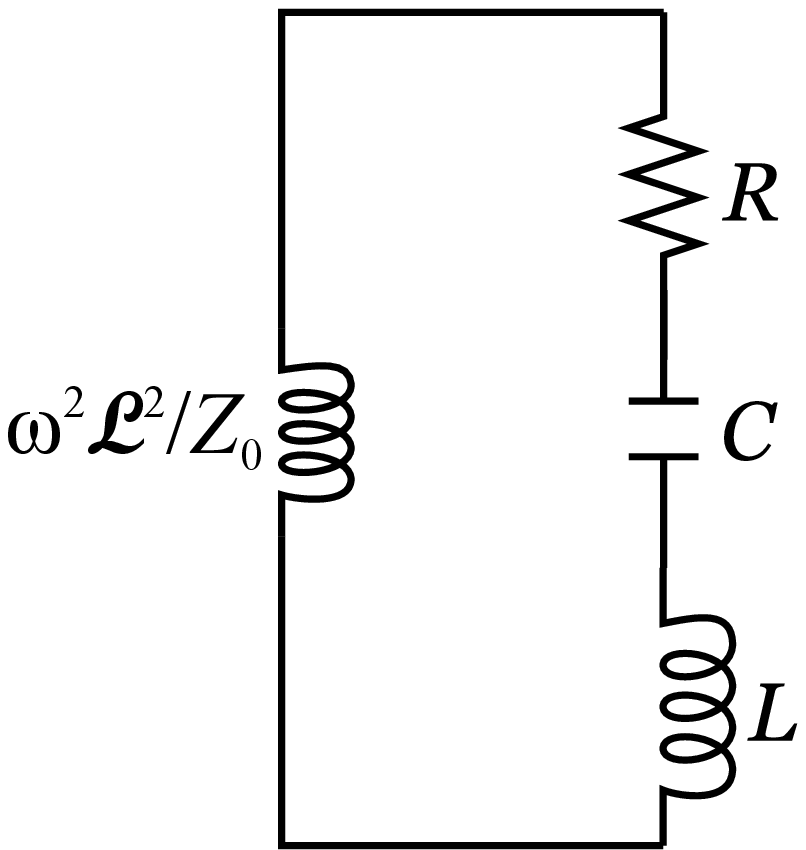}
\caption{Equivalent RLC-circuit of iris coupled high-$Q$ cavity near resonance\cite{princmic}.}
\label{fig:rlc-iris}
\end{minipage}
\end{figure}

The integro-differential equation for the discharge transient reads:
\begin{equation}
0=\frac{1}{C}\int I \, \dd t + (L + \mathcal{L})\frac{\dd I}{\dd t} + \left(R + \frac{\omega^2 \mathcal{L}^2}{Z_0} \right) I,
\end{equation}
After differentiating a second order differential equation is obtained:
\begin{equation}
0=\frac{I}{C} + (L + \mathcal{L}) \frac{\dd^2 I}{\dd t^2} + \left(R + \frac{\omega^2 \mathcal{L}^2}{Z_0} \right) \frac{\dd I}{\dd t}.
\end{equation}

Using the $I(t)=A\expe^{-\lambda t}$ ansatz, the characteristic equation near resonance is:
\begin{equation}\label{eq:chareq}
\lambda_{1,2}= \frac{R + \frac{\omega_0^2 \mathcal{L}^2}{Z_0}}{2(L+\mathcal{L})} \pm \cplxi \sqrt{ \left(\frac{R + \frac{\omega_0^2 \mathcal{L}^2}{Z_0}}{2(L+\mathcal{L})} \right)^2 - \frac{1}{C(L+\mathcal{L})} }.
\end{equation}
Here, critical coupling is achieved when
\begin{equation}
\mathcal{L}^2 \omega_0^2 = Z_0 R.
\end{equation}

According to Ref. \onlinecite{princmic} the following equalities hold between the circuit model terms and the microwave cavity parameters:
\begin{gather}
\omega_0 = \frac{1}{\sqrt{C(L+\mathcal{L})}}, \\
\frac{\omega_0}{Q_0} = \frac{R}{L+\mathcal{L}}, \\
\frac{\omega_0}{Q_{\text{c}}} = \frac{\omega_0^2 \mathcal{L}^2}{Z_0(L+\mathcal{L})}, \\
Q^{-1} = Q_0^{-1} + Q_{\text{c}}^{-1}, \\
Q = Q_0/(1+\beta),
\end{gather}
With these, Eq. \eqref{eq:chareq} becomes
\begin{equation}\label{eq:lambda}
\lambda_{1,2} = \frac{\omega_0}{2 Q} \pm \cplxi \sqrt{\omega_0^2 - \left( \frac{\omega_0}{2 Q} \right)^2}.
\end{equation}

The general solution for the discharge transient is
\begin{equation}
I(t) = \sum\limits_{i=1}^2 A_i \expe^{-\lambda_i t},
\end{equation}
where the $A_i$ coefficients can be fitted to satisfy the initial conditions.
It is emphasized that according to Eq. \eqref{eq:lambda} the transient has a characteristic decay rate
of $\omega_0 / 2Q$, which is in an agreement with the previous calculations and with Eqs. (1) and (2) in the main paper. The frequency of the oscillations is the eigenfrequency of the coupled circuit (i.e. it contains a small shift due to the coupling).

\section{Limit of detection for significant detuning}

We first derive Eq. (3) in the main paper. The energy stored inside a critically coupled microwave cavity has a maximum of $U_{\max}=p_0 Q_0/\omega_0$ for $f=f_0$ and it falls as a function of the microwave frequency detuning, $\omega-\omega_0$ with a Lorentzian of
\begin{equation}
U(\omega-\omega_0)=U_{\max}\frac{\left(\frac{\Delta \omega}{2} \right)^2}{\left(\frac{\Delta \omega}{2} \right)^2+\left(\omega-\omega_0\right)^2}.
\label{eq:stored_energy_detuned}
\end{equation}
It was discussed above that the emitted microwave energy during the switch off transient is half of the total stored energy. Therefore Eq. \eqref{eq:stored_energy_detuned} yields Eq. (3) in the main paper for the emitted energy:
\begin{equation}
U_{\text{em}}(\omega-\omega_0)=\frac{p_0 Q_0}{2 \omega_0}\times\frac{\left(\frac{\Delta \omega}{2}\right)^2}{\left(\frac{\Delta \omega}{2}\right)^2+\left(\omega-\omega_0\right)^2}.
\end{equation}

The emitted energy is the signal strength of the measurement. The energy of the noise is
\begin{equation}
U_{\text{noise}} = 4 k_{\text{B}} T \times \mathrm{ENBW}/1~\text{Hz},
\end{equation}
where $\mathrm{ENBW}$ stands for Equivalent Noise Bandwidth and the division by $1$ Hz is required to match the dimensionality. The factor $4$ appears due to the well known $6$ dB noise figure of the microwave mixers. The $\mathrm{ENBW}$ of a Fourier transformed signal with a rectangular apodization is known to be reduced by the number of points, $N$, of the measurement bandwidth, $\mathrm{BW}$, and reads:

\begin{equation}
\mathrm{ENBW} = \frac{\mathrm{BW}}{\sqrt{N}}.
\end{equation}

We assume that a well designed experiment can be performed, i.e. the measurement bandwidth exceeds the detuning frequency and that the number of points can be selected such that ENBW is about 10 times smaller than $\Delta \omega$.

Considering the strongly detuned case, $\Delta \omega \ll \omega - \omega_0 \equiv \omega_{\text{d}}$ and substituting the definition of $Q_0$ one can gain that
\begin{equation}
U_{\text{em}}(\omega_{\text{d}}) = \frac{p_0 \Delta \omega}{4 \omega_{\text{d}}^2}.
\end{equation}

The Signal/Noise ratio is in our case then reads
\begin{equation}
\mathrm{S}/\mathrm{N} = U_{\text{em}}(\omega_{\text{d}})/U_{\text{noise}} = \frac{10 p_0\times 1\,\text{Hz}}{16\omega_{\text{d}}^2 k_{\text{B}}T}.
\label{eq:S_per_N}
\end{equation}
Eq. \eqref{eq:S_per_N} is remarkably parameter independent and it contains only the detuning frequency, the microwave power, and the thermal noise.

Taking $p_0 = 1~\text{mW} = 0~\text{dBm}$ the maximal detuning becomes
\begin{equation}
\omega_{\text{d}}/2\pi \approx 60~\text{MHz},
\end{equation}
in agreement with our measurements. We note that with the use of an increased irradiation power or a cooled microwave mixer, detuning frequencies of a few GHz could be achieved.

\section{Additional cavity transient data for the undercoupled case}

\begin{figure}[h!]
\begin{center}
\includegraphics[width=0.500\textwidth]{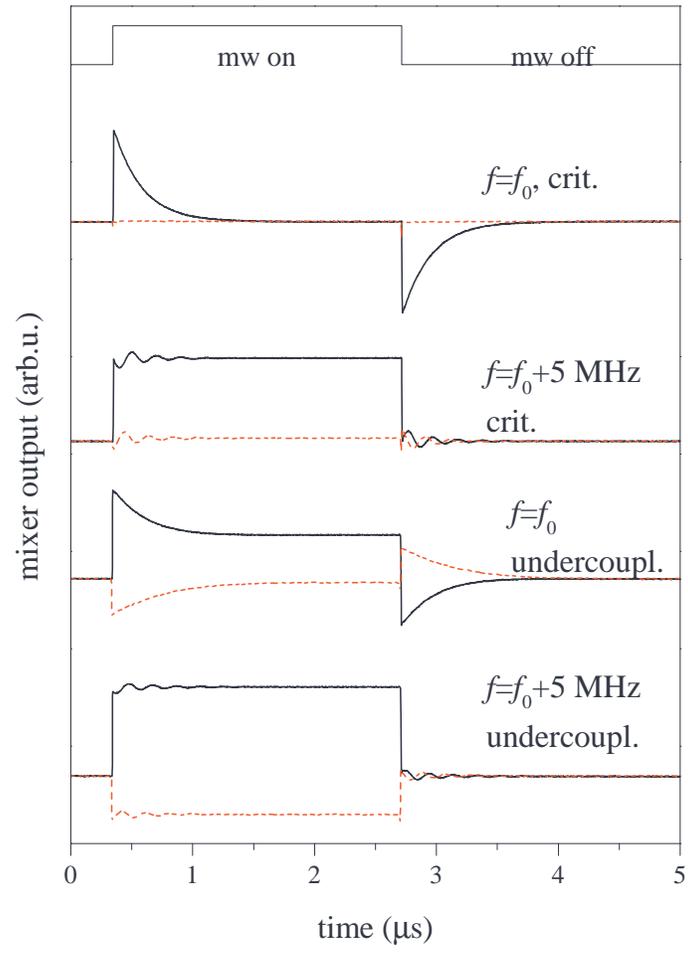}
\caption{Time domain signals shown for critical and undercoupled cases. For the latter, half of the exciting power is reflected from the cavity when the irradiation is on resonance and the steady state is achieved.}
\label{fig:TimeDom_trans_undercoupl}
\end{center}
\end{figure}

In Fig. \ref{fig:TimeDom_trans_undercoupl}. we show the time domain signals for both the critically and undercoupled cases. For the latter, the coupling is such that half of the exciting power is reflected from the cavity. The critically coupled data is the same as that shown in Fig. 2. of the main article.

\begin{figure}[h!]
\begin{center}
\includegraphics[width=0.500\textwidth]{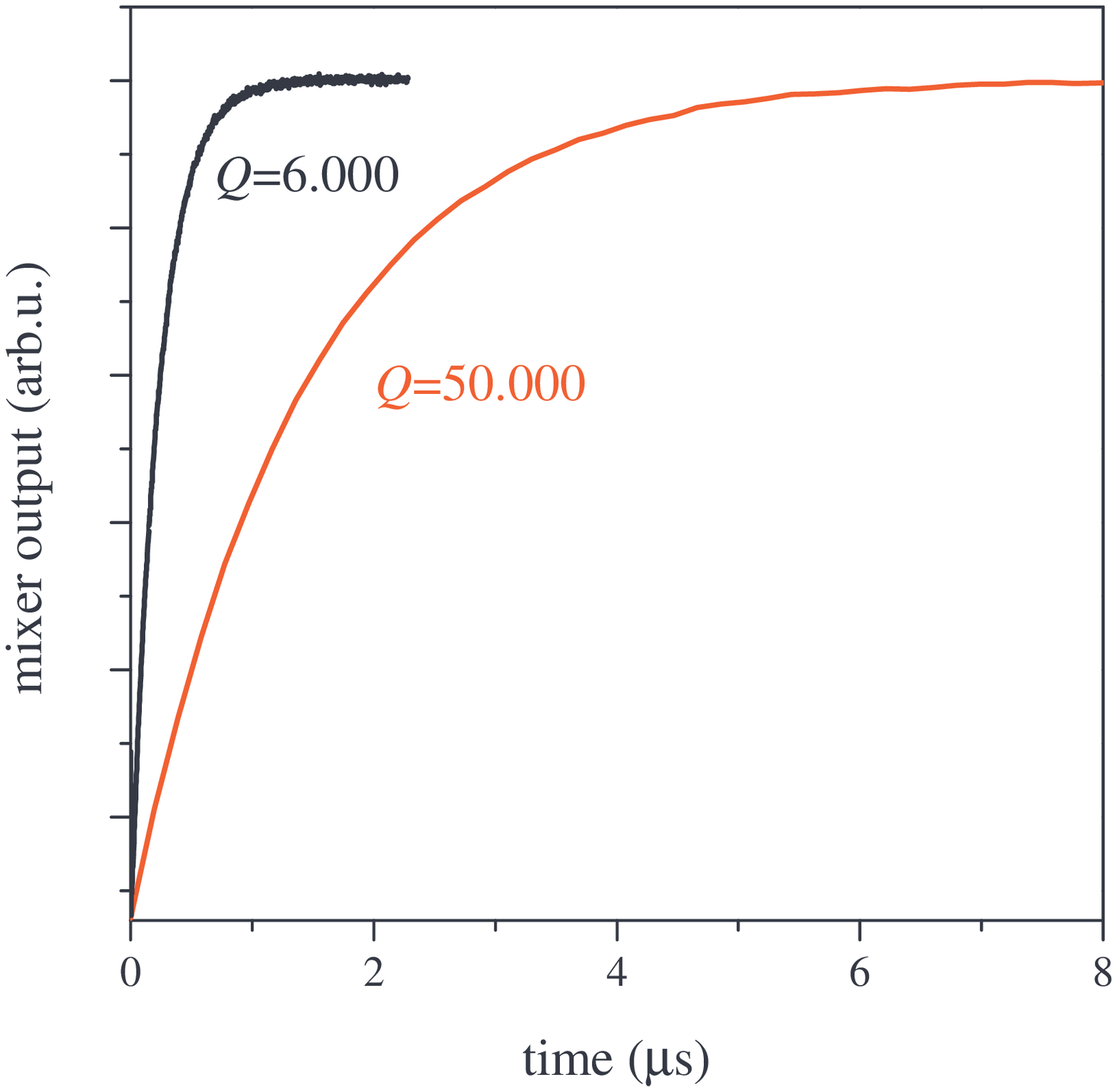}
\caption{Comparison of the cavity transient signal for two cavities with different $Q$'s of copper ($Q\sim 6.000$) and niobium at 4.2 K ($Q \sim 50.000$). Both cavities were critically coupled and the out-of-phase quadrature mixer is nulled (data not shown).}
\label{fig:HighQ_lowQ_compare}
\end{center}
\end{figure}

In Fig. \ref{fig:HighQ_lowQ_compare}., we compare the cavity transient signals for two different cavities of copper ($Q\sim 6.000$) and niobium at $4.2$ K ($Q \sim 50.000$). Superconducting cavities can reach $Q$ factors beyond one million but our niobium had an off the self purity.



\end{widetext}

\end{document}